\documentclass[runningheads]{llncs}
\usepackage{graphicx}
\usepackage[utf8]{inputenc} 
\usepackage[T1]{fontenc}    
\usepackage{hyperref}       
\usepackage{url}            
\usepackage{booktabs}       
\usepackage{amsfonts}       
\usepackage{nicefrac}       
\usepackage{microtype}      
\usepackage{lipsum}
\usepackage{amsmath,amssymb}
\usepackage[ruled,vlined]{algorithm2e}
\usepackage{rotating}
\DeclareMathOperator{\E}{\mathbb{E}}

\begin{document}
\title{Fundus Image Analysis for Age Related Macular Degeneration: ADAM-2020 Challenge Report }
\titlerunning{ADAM-2020 Challenge Report}
%
\author{Sharath M Shankaranarayana}
%
\institute{Zasti.AI\\
\email{sharath@zasti.ai}}
%
\maketitle              
\begin{abstract}
Age related macular degeneration (AMD) is one of the major causes for blindness in the elderly population. In this report, we propose deep learning based methods for retinal analysis using color fundus images for computer aided diagnosis of AMD. We leverage the recent state of the art deep networks for building a single fundus image based AMD classification pipeline. We also propose methods for the other directly relevant and auxiliary tasks such as lesions detection and segmentation, fovea detection and optic disc segmentation. We propose the use of generative adversarial networks (GANs) for the tasks of segmentation and detection. We also propose a novel method of fovea detection using GANs.
\end{abstract}
\section{Introduction}
With the advancement in medical field and thereby increase in life expectancy, age-related diseases also tend to increase, which causes burden on the healthcare providers. Age related macular degeneration (AMD) is one such diseases which affects the elderly and potentially causing loss in vision. Early detection is very important for prevention and treatment of AMD. Color fundus imaging (CFI) is one of the quickest retinal imaging modality and is very useful in monitoring a large number of retinal diseases. However, conclusive diagnosis for AMD is done predominantly by examining other retinal imaging modality called optical coherence tomography (OCT), since most ophthalmologists find it difficult to accurately diagnose AMD only based on CFI. Moreover, the task of detecting the abnormalities in retina such as drusen, exudate, hemorrhage etc., is a labor intensive and time consuming process even for experts. This necessitates the need for automated methods for fundus image analysis for detection of AMD.

There are only a few works on automated AMD detection using color fundus images. Recently the authors of \cite{deepseenet} proposed a deep convolutional neural network (CNN) based classification system for AMD along with providing a large dataset for AMD related research. Instead of directly classifying retinal images as AMD or not, the authors use Age-Related Eye Disease Study (AREDS) Simplified Severity Scale to predict the risk of progression to late AMD. In another recent work \cite{wang2019two}, the authors propose a deep network which detects the presence of dry and wet AMD using both fundus images as well as OCT slices.

In this report, we propose methods employed for various tasks related to retinal image analysis for aiding the detection of AMD. We propose methods for single image level grading for AMD using CFI and also methods for segmentation of various kinds of lesions found in the retinal. We also propose a novel method for the localization of fovea in the fundus images.


\section{Methodology}
\subsection{Task 1: AMD Classification}
The first of the challenge requires us to predict the probability of AMD for a given retinal fundus image. The dataset consists of $400$ training images of which $89$ have AMD and the rest $311$ do not have AMD. To overcome the inherent imbalance in the dataset, we resort to  data augmentation along with oversampling of the AMD fundus images. For data augmentation, we employ the following techniques:
\begin{enumerate}
    \item Random flipping and rotation
    \item Photometric distortion
    \item Specific histogram based image processing techniques such as histogram equalization, adaptive histogram equalization, intensity rescaling at different levels, histogram matching by randomly selecting a few cannonical images from the validation set.  
\end{enumerate}
\begin{figure}[tp]
   \centering
\includegraphics[width=\textwidth]{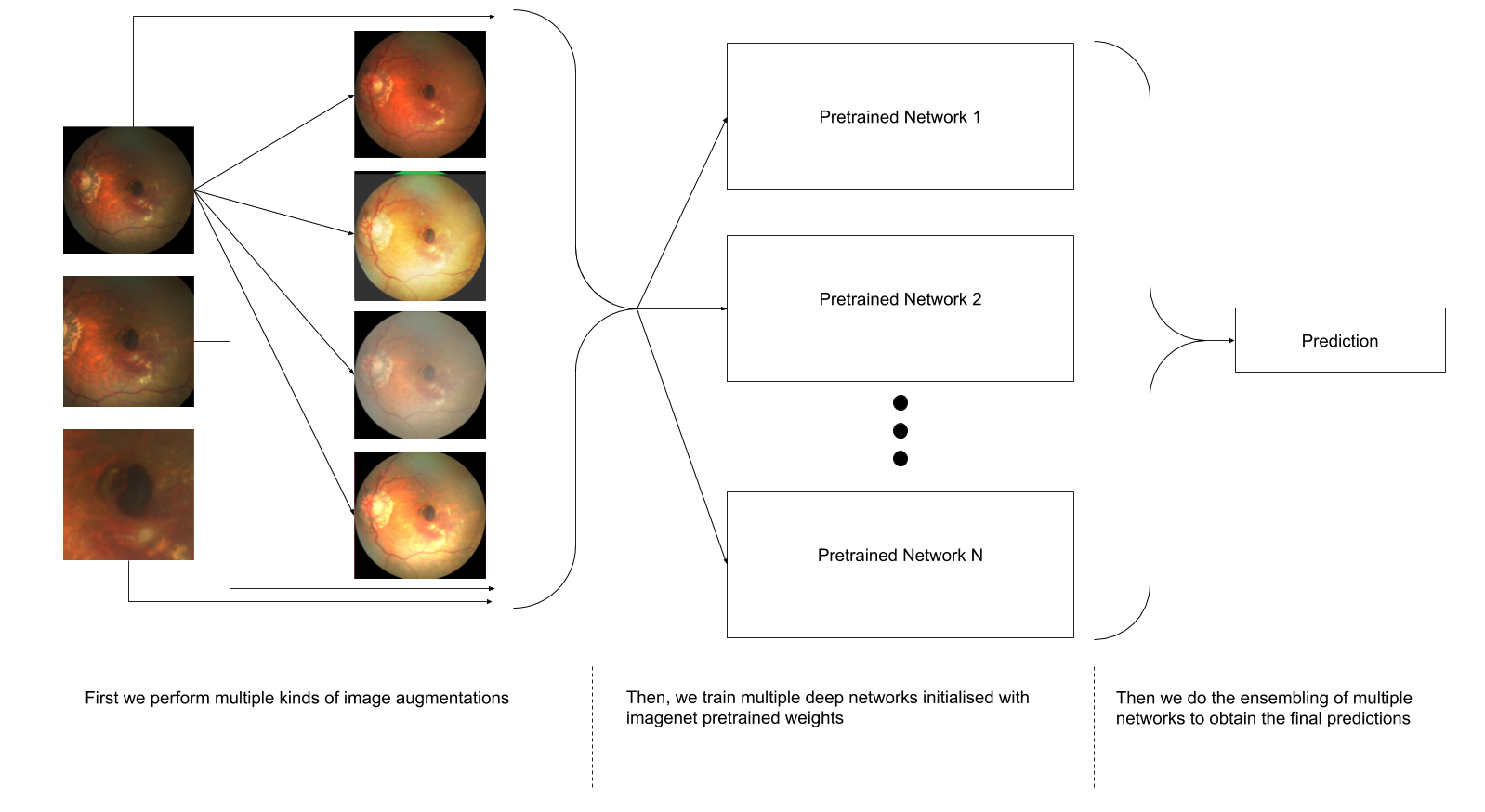} 
         \caption{Proposed pipeline for AMD classification}
  \label{FigT1}
   \end{figure}

With the augmented dataset, we then employ multiple pretrained deep convolutional neural networks (CNNs) for the task of binary classification. The classification networks employed are: 
\begin{enumerate}
    \item EfficientNet \cite{efficientnet}: EfficientNets are recently introduced class of networks and they employ a model scaling method to scale up CNNs in a more structured way and they have been shown to surpass the performance of other deep networks on ImageNet dataset and with better efficiency. We use multiple classes of EfficientNets- EfficientNet-B4, EfficientNet-B5, EfficientNet-B6, EfficientNet-B7. We use them because they are optimized for training at larger resolutions ($380$, $456$, $528$ and $600$, respectively) when compared to standard $224$x$224$ resolution of other Imagenet pretrained networks.    
    
    \item Inception-Resnet \cite{inception_resnet}: We use the Inception-Resnet architecture since it combines the two most commonly used blocks- inception block, which helps in multiscale feature extraction and residual block, which helps in faster convergence and alleviating vanishing gradients. These two blocks help in better feature extraction. 
    
    \item Resnext \cite{Resnext}: Resnext is another highly modularized network architecture for image classification tasks, which has also proved to be state-of-the-art in ImageNet classification task. Along with ImageNet pretrained Resnext, we also use pretrained weights obtained by weakly supervised learning on the Instagram dataset \cite{wsl}. 
    
    \item Squeeze and Excitation networks \cite{senet}: We use this  class of network architectures since they consist of "Squeeze-and-Excitation" (SE) blocks that tend to generalize well across different datasets.   
\end{enumerate}
Since AMD is characterized by abnormalities in macular region, we also crop out the macular region at various different zoom levels. Finally, with all the networks trained for the AMD classification task, we ensemble the network predictions using simple averaging of posterior probabilities (overall block diagram shown in figure Fig \ref{FigT1}.

\subsection{Task 2: Optic Disc Segmentation}

For the second task, we train a deep network for the task of semantic segmentation of OD. We employ the same methodology proposed in our previous works \cite{resunet} for training. But instead of using the RGB color channels, we use inverted green channel images as proposed in \cite{jbhi} since the inverted green channel images provides better contrast for OD and the background. We use adversarial training setting and use ResUnet as our base architecture as mentioned in the paper \cite{resunet}. The reader is advised to refer \cite{resunet} for details. Once we train the network and predict the segmentation maps for the given retinal images, we perform some post-processing operations on the predicted maps. We first keep only the largest connected component in the binary segmentation map and remove the other smaller components. And later we apply convex-hull operator to obtain the final segmentation map. 

Additionally, the task also requires us to do optic disc detection. For this, we simply keep a threshold on the area largest connected component in the segmentation map during the post-processing stage and discard if the area is smaller than the threshold.      
\begin{figure}[tp]
   \centering
\includegraphics[width=\textwidth]{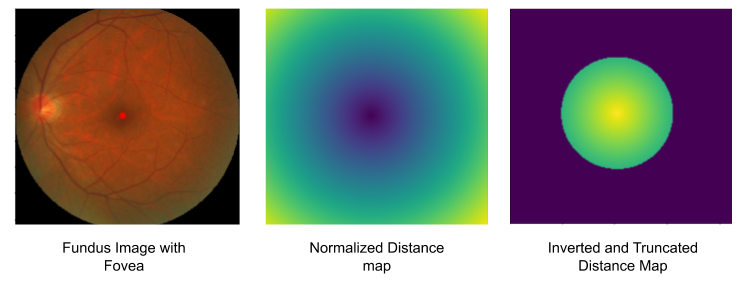} 
         \caption{Process of creating distance maps}
  \label{Dmaps}
   \end{figure}
\subsection{Task 3: Fovea Localization}
\begin{figure}[tp]
   \centering
\includegraphics[width=\textwidth]{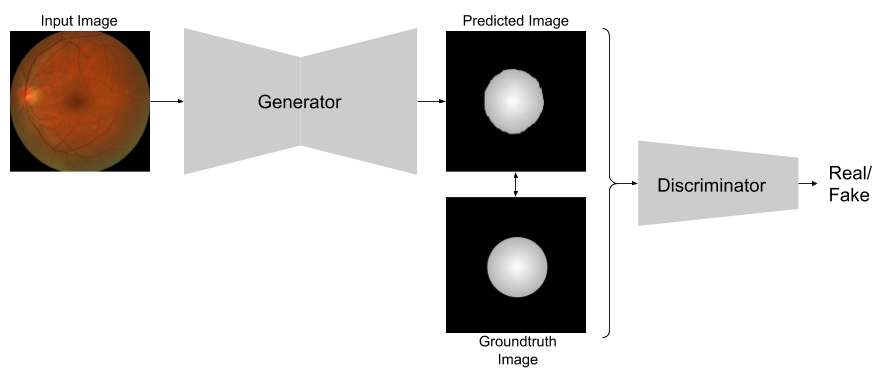} 
         \caption{Overall block diagram of fovea distance map regression}
  \label{FigT3}
   \end{figure}

This task requires us to predict the point coordinates of the location of fovea. Instead of treating it as a standard coordinates regression problem, we convert this into an image to image translation problem. For this, prom the information of groundtruth point coordinates, we create distance maps having the same size as the dundus images. Distance maps are using the euclidean distance transforms computed from the fovea location i.e., the distances farther from the fovea have higher values than the distances closer to the fovea. For the purpose of easier training, we normalize the distance map and then invert so that the distances nearer to fovea has higher values. We later truncate it such that it only contains a specific radius around the fovea. This is done to improve training and is shown in figure Fig. \ref{Dmaps}.We then treat it as an paired image to image translation problem \cite{unpaired} as we did with od segmentation task similar to our work in \cite{resunet}. The overall block diagram of generative adversarial network (GAN) framework used for image translation task is shown in figure Fig.  \ref{FigT3}

We train the generator is to learn a mapping from the input fundus image $x$ to the fovea distance map $y: G : x  \rightarrow y$. We train the discriminator to distinguish between the generated distance map and real distance map: 
\begin{equation}\label{eq2.1}
L_{GAN}(G,D)=\E_{x,y}[log(D(y))] +
               \E_{x}
               [log (1-D(G(x))]
\end{equation}

where $\E_{x,y}$ represents the expectation of the log-likelihood of the pair $(x,y)$ being sampled from the underlying probability distribution of real pairs of input fundus images and groundtruth distance maps.

Additionally, we also use $L_1$ loss between the generator predicted distance map and groundtruth distance map and therefore the final objective becomes
\begin{equation}\label{eq4}
G^* = arg \underset{G}{\mathrm{min}} \underset{D}{\mathrm{max}}( L_{GAN}(G,D) + \lambda(L_{L1}(G))
\end{equation}
where $\lambda$ balances the contribution of two losses. In equation (\ref{eq4}), the discriminator tries to maximize the expression by classifying whether the distance map is real or generated. The generator tries to minimize both adversarial loss and $L1$ loss in equation (\ref{eq4}).

The architecture for generator part of GAN is described in the next section since the base architecture is same for this task as well as the lesions segmentation task. The discriminator uses a conventional CNN architecture used for classification.  Let Ln denote  a  Convolution-BatchNorm-ReLU  layer with n filters. The discriminator uses the following architecture \\
L64-L128-L256-L512-L512-L512 \\

Finally, with the predicted distance maps from the generator, we extract the fovea point coordinates by performing a post-processing operation. Ideally, we could just take the pixel with highest intensity as the fovea coordinate, but doing so results in other erroneous regions as well. Therefore we cluster one percent of the highest intensities and segment out the largest cluster. The centroid of this largest cluster gives us the fovea coordinate.

\subsection{Task 4: Lesions Segmentation}

For this task, given a fundus image, we need to segment out various kinds of lesions such as drusen, exudate, hemorrhage, scar and others. Similar to the previous two tasks, we employ GAN based frameworks for the task of lesions segmentation. The only difference between OD segmentation in Task 2 and this task is the generator architecture. We would also like to mention once again that this generator architecture was used for the fovea localization task as well. 
\begin{figure}[tp]
   \centering
\includegraphics[width=\textwidth]{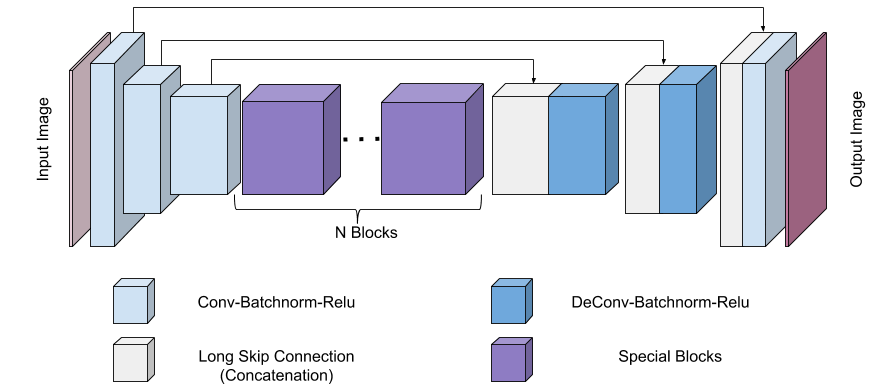} 
  \caption{Base Architecture or Generator Archtecture for GAN}
  \label{Fig. MainArchitecture}
\end{figure}
\subsubsection{Base Architecture :}The architecture for the base network (or the generator network is shown in Fig. \ref{Fig. MainArchitecture}. The input input image is first passed through three convolutional layers with kernel sizes $7X7$, $3X3$ and $3X3$ respectively. We use strided convolution with stride of 2 for downsampling the second and third layers. These initial layers serve as coarse feature extraction layers. Then, the network consists of special blocks which are inspired from \cite{resnet} and \cite{resunet} works.  Then, after specific number of special blocks, the network consists of strided deconvolutional layers for upsampling and finally a $1X1$ convolution for mapping to the output. All the layers are followed by batch normalization and $Relu$ operations, except the final layer which is followed by $tanh$ activation. Additionally, we also employ long skip connections.     \\

We train the GAN for each of the lesion segmentation task separately, since there is severe class imbalance and also overlap of the lesion annotations in some cases. Thus, we predict segmentation maps separately for each of the tasks. Finally, for lesions detection, we simply discard those segmentation predictions where the lesion area is less than a specific threshold value found empirically.

\section{Experiments and Results}

For the first task of AMD classification, we employ only the images provided by the challenge organisers for training, and a subset of STARE dataset \cite{stare} for validation. We train a total of $7$ models with $3$ different crop levels with respect to macular region. We thus obtain $21$ different models. All the networks are trained for $20$ epochs and stochastic gradient descent (SGD) optimizer except the EfficientNets, which are trained using ADAM optimizer, with an initial learning rate of $1e^{-3}$ for all models. We save the model giving the highest accuracy on the validation set. While testing, we perform test time augmentation (TTA) by doing various histogram processing based operations for each of the trained networks and finally perform ensembling of all models by taking the mean of posterior probabilities for AMD. The final ensembling gave us an AUC score of $0.957$ in the validation stage of the competetion.    

For the second and third tasks, along with the challenge data, we employ $1200$ images from the REFUGE challenge \cite{refuge} , while for the fourth task, we only use the dataset provided by the organizers of the competetion. We train the GAN models from scratch with initialization  from  a  Gaussian  distribution  with  mean  0  and standard deviation 0.02 and do the training for $200$ epochs for the second and the third tasks since we have sufficient number of images after augmentation. Since the number of images for the fourth task are less, even after extensive augmentation, we initialize the weights obtained from the fovea distance map regression task. This is also helpful since most of the abnormalities occur near the macula region than the OD region.  We keep the initial learning rate to be $10^{-4}$ and halve every $50$ epochs. We train the models at a high resolution of $640$x$640$. We save the model which gives the best validation score. For the second task, we obtain an $F1$ score of $1.0$ for disc detection, and $Dice$ score of $0.9653$ for OD segmentation. For the third task, we $Euclidean$ $Distance$ difference of $25.589$ pixels. We obtain $Dice$ scores of $0.5381$, $0.4383$, $0.4148$, $0.5446$ and $0.2628$ for the tasks of drusen, exudate, hemorrhage, scar and others segmentation respectively.

\section{Conclusion}
In this report, we outlined the methods for fundus image analysis for various AMD related tasks. All of our techniques employed the latest advancements in deep learning. We see that even directly employing the latest classification networks off the shelf provides a good initial score. Also, the GAN based models perform very well in almost all of the other tasks. In future, we would like to explore multi-task learning for obtaining better results.

\section{Acknowledgement}
We thank the organizers of the Automatic Detection challenge on Age-related Macular degeneration (ADAM) (https://amd.grand-challenge.org/) for hosting the challenge and kindly providing the dataset. 

%
%
\bibliographystyle{splncs04}
\bibliography{mybibliography}
%




\end{document}